\documentclass[aps,prl,preprint,superscriptaddress]{revtex4-1}

\usepackage{amsfonts}
\usepackage{graphicx}
\usepackage{epsfig}
\usepackage{epsf}
\usepackage{amssymb}
\usepackage{amsmath}
\usepackage{amsthm}
\usepackage{multirow}
\usepackage[table,xcdraw]{xcolor}
\usepackage{mathrsfs}
\usepackage{bm}
\usepackage{times}
\usepackage{hyperref}
\usepackage{epstopdf}
\hypersetup{colorlinks,
	linkcolor=black,
	filecolor=black,
	urlcolor=blue,
	citecolor=blue,}

\newcounter{lastnote}

\begin{document}
\title{Non-reciprocal Radio Frequency Transduction in a Parametric Mechanical Artificial Lattice}

\author{Pu Huang}
\thanks{These authors contributed equally to this paper and are joint first
	authors.}
\author{Liang Zhang}
\thanks{These authors contributed equally to this paper and are joint first
	authors.}
\affiliation{Hefei National Laboratory for Physical Sciences at Microscale and Department of Modern Physics, University of Science and Technology of China, Hefei 230026, People's Republic of China}
\affiliation{Synergetic Innovation Center of Quantum Information and Quantum Physics, University of Science and Technology of China, Hefei 230026, People's Republic of China}
\author{Jingwei Zhou}
\author{Tian Tian}
\author{Peiran Yin}
\affiliation{Hefei National Laboratory for Physical Sciences at Microscale and Department of Modern Physics, University of Science and Technology of China, Hefei 230026, People's Republic of China}
\author{Changkui Duan}
\author{Jiangfeng Du}
\thanks{Corresponding author: \href{mailto:djf@ustc.edu.cn}{djf@ustc.edu.cn}}
\affiliation{Hefei National Laboratory for Physical Sciences at Microscale and Department of Modern Physics, University of Science and Technology of China, Hefei 230026, People's Republic of China}
\affiliation{Synergetic Innovation Center of Quantum Information and Quantum Physics, University of Science and Technology of China, Hefei 230026, People's Republic of China}

\date{\today}


\begin{abstract}
Generating non-reciprocal radio frequency transduction plays important roles in a wide range of research and applications, and an aspiration is to integrate this functionality into micro-circuit without introducing magnetic field, which, however, remains challenging. By designing a 1D artificial lattice structure with neighbor-interaction engineered parametrically, we predicted a non-reciprocity transduction with giant unidirectionality. We then experimentally demonstrated the phenomenon on a nano-electromechanical chip fabricated by conventional complementary metal-silicon processing. A unidirectionality with isolation as high as $24~$dB is achieved and several different transduction schemes are realized by programming the control voltages topology. Apart from being used as a radio frequency isolator, the system provides a way to build practical on-chip programmable device for many researches and applications in radio frequency domain.
\end{abstract}

\maketitle



In crystal lattice of normal matter, the behavior of mechanical oscillating propagation follows a general symmetry known as reciprocity that makes wave transport bidirectional, and such a property originates from the time-reversal symmetric nature of the micro-scale chemical interactions between the atoms of the lattice. Familiar examples are reversibility of optical and acoustic wave \cite{Landau1986em,Landau1986fm} encountered in our everyday life. To generate unidirectional transport for specific applications, a direct way is to break the time reversal symmetry, and in electro-magnetic domain,
special magnetic effects are explored and widely used for such a purpose, such as Faraday rotation \cite{NPHO2011}. However, it becomes awkward when the frequency goes from optical to radio frequency (RF) domain, where a lot of applications exist such as building classical and quantum communication networks and weak signal measurement. This is because the extending of wavelength into the scale of meters makes the corresponding unidirectional component bulky, while active methods, such as using transistors, bring about additional noise and high power consumption \cite{TE1965,APL2011} and so become incompatible for high performance and integration purposes. Therefore, realization of chip-scale RF unidirectional component without these drawbacks is extremely valuable.


Extensive efforts have been paid to exploring new way to magnetic-free unidirectional transduction recently. A passive way is to make use of the nonlinear response of the medium propagation \cite{JAP1994,JOS2002,NM2005,SC2012}, and other active ways include modulating the medium by applying an external field that works effectively as magnetism of the medium \cite{NPHO2009,NP2011,NPHO2014,PRX2015,NP2015}, and engineering the optical cavity's internal symmetry  \cite{NPHO2014_2,NP2014}.
These schemes have characteristic sizes limited by the wavelength of the wave.
A subtle scheme that exploring the medium's angular momentum is  proposed \cite{NC2013}, which well releases the limitation on the device by wavelength
and has been experimentally explored both by using acoustic wave in fluid and by electric circuit \cite{SC2014,NP2014_2}. Inspired by topological states in condensed matter,
unidirectional transports of classic waves as topological edge states have been observed  \cite{NA2009,NM2012,SC2015}, but the constructed devices for RF domain are far from integration.


In an artificial lattice, the micro-scale interaction between ``atom'' is no more given by chemical interaction but can be engineered externally \cite{SC2015,NA2002,SC2011,NA2013}, therefore special behaviors different from those in normal crystal lattices can be generate. By introducing a time-dependent interaction which breaks time-reversal symmetry, non-reciprocity is expectable. For practical realization at radio frequency domain, a system of nano-electromechanical components serves as an ideal candidate due to its natural RF frequency matching and  readiness for integration puropose \cite{SC2000,RSI2005}. Besides, its compatibility to conventional CMOS processing, convenient interface to integrated micro-circuit, as well as its ability to function in ambient condition, make it promising for various further practical applications. In this letter, we demonstrated such an parametric artificial lattice using a radio-frequency nano-mechanical chip, giant non-reciprocally transduction behaviour are generated and the controllability in the parametric interaction make the structure configurable for different transduction scheme.



The model system we constructed contains a chain of 1D harmonic oscillators as schematic in Fig.~1. Each oscillator of vibrational amplitude $x_i$ couples to adjacent ones parametrically with spatial and time dependent strength described by the following equation:
\begin{equation}
\label{EOM} m_i\ddot{x}_{ i}  =-[m_i\omega_{ i}^2 {\bm \delta}_{i,j} -{\bm L}_{i,j}(t)] x_{j},
\end{equation}
where $\omega_i$ and $m_{i}$ are the circular frequency and the mass of oscillator $i$, respectively, and ${\bm L}_{ij} (t)$ is time-dependent parametric coupling matrix with its nonzero elements satisfying ${\bm L}_{i,i+1} (t) ={\bm L}_{i+1,i} (t) =-\eta_{i} {\cos}(\omega^{p}_{i} t)\delta_{i}(t)$, and  ${\bm L}_{i,i} (t)= -[{\bm L}_{i,i+1} (t)+{\bm L}_{i,i-1} (t)]$, with $\eta_{i}$ the coupling strength, $\omega_i^{ p}$ the pump frequency and $\delta_i(t)$ the modulation function.  The pump frequency is chosen to satisfy the frequency conversion condition \cite{PRL2013}
\begin{equation}
\label{PFC} \omega^{p}_{ i}= |\omega_{ i}-\omega_{ i+1}|,
\end{equation}
so as to convert the motions between oscillators $i$ and $i+1$. The modulation takes the form
\begin{equation}
\label{Mod} \delta_{i}(t)= \cos{  (\Omega t - i\Theta)},
\end{equation}
with $\Omega$ and $\Theta$ the modulation frequency and delay phase, respectively. A time delay is then defined as $\tau=\Theta/\Omega$. The key difference of such an artificial lattice from a natural chemical crystal lattice lies in the coupling matrix ${\bm L}_{i,j} (t)$, whose time-reversal symmetry (T-symmetry) is apparently broken since ${\bm L}_{i,j}(-t) \neq {\bm L}_{i,j}(t)$. However, when time reversal and space reflection transformation (P-symmetry) are taken simultaneously, the equation of motion is invariant, showing the property of PT-symmetry (see SM for detail).

In real physics, apart from the dynamics dictated by Eq.\ (\ref{EOM}), a universal dissipation that leads to the transfer of coherent energy of the system into environment needs to be included. As a result, when a resonant external excitation is applied onto the leftmost (or rightmost) oscillator, the energy transports coherently to the rightmost (or leftmost) one with only weak loss for the intended transduction direction, but decays almost entirely to the environment for the reversed direction. Both the transduction direction and the isolation (defined below) can be controlled by setting suitable parameters. Specifically, by setting the  modulation as constant, i.e., $\delta_{i}(t) =1 $, the system behaviors as a common bidirectional transmission line,  but when the modulation is tuned to be positive, $\Theta>0$ (negative delay, $\Theta<0$), the transduction from left to right is allowed (suppressed) but the reversed is suppressed (allowed), as shown in Fig.~1(d). The interaction described above does not bring in any additional noise \cite{PRL2013}, as can be seen straightforwardly by going to the quantum mechanics picture, where the parametric condition of Eq.\ (\ref{PFC}) ensures the conservation of the system's total phonon numbers, which, in the limit of  $\omega_{i} = \omega_{j} $, indicates the conservation of total energy and therefore passive (see SM for detail).


To further understand the system, we numerically solved Eq.\ (\ref{EOM}) with dissipation (see the model in SM).  The parameters for lattice with $N$ oscillators are set to be symmetric under space reflection in absence of parametric coupling, i.e. when ${\bm L}_{i,j} (t)=0$. A resonant excitation force $F_{i}(t)=F \cos(\omega_{i}t)$  is applied on oscillator $i$ ($i =1$ or $N$), then the amplitude of vibration $x_{j}$ ($j=N$ or $1$) is registered after the system becomes stationary. The unidirectionality of forward transduction is characterized by the forward isolation $ {{I}_{\rm for}} = (x_{1}/x_{N})^2$. Correspondingly, the backward isolation is defined as ${I_{\rm back}}= (x_{N}/x_{1})^2$.

Fig.~2(a) plots typical forward isolation values ${{I}_{\rm for}}$ as a function of the coupling strength $\eta$ (by setting all $\eta_{i} = \eta $) with corresponding normalized coherence length $L_c$ (see SM for definition) and the modulation frequency $\Omega$. In the simulation, the system's lattice number $N = 9$ and a positive phase delay $\Theta = \pi/8$ is adopted. The results show that, when the coherent length is much smaller than the lattice size ($L_{c}\ll 1$), the unidirectional behaviors are insensitive to $\eta$ but depend strongly and especially on the sign of $\Omega$, indicating one of the motion transduction decaying predominantly into the environment. While when $L_c$ become comparable to the lattice size ($L_{c} \geq 1$), complicated behaviors emerge, indicating that the motion propagation reflects many times in the lattice before decaying into environment. On the other hand, as shown in Fig.~2(b), space reflection, as expected, can also change the system's behaviors.

In practical usage,  in additional to the isolation, the transduction loss is also of concern. We can tuning both the coupling strength and $\Omega$ so that one of transduction is strongly decayed while maintaining the other one as only weakly decayed. This is realized, as shown in Fig.~2(c),  by increasing $N$: the forward transduction loss is maintained to about $-6~$dB for all $N$, while the isolation is enhanced. Fig.~2(d) plots the frequency response for $N = 9$, which shows, as expected, that the system suppresses backward transduction (${{I}_{\rm for}} > 1 $). Fig.~2(e) and 2(f) together with Fig.~2(d) show that the transduction topology can be easily controlled by changing modulations. This makes such an artificial system programmable for signal transduction.


We demonstrated the model by using an experimental system composed of a chain of doubly clamped beams made on a SOI wafer with standard nano-fabrication process. The picture of the device's chip is shown in Fig.~3(a), and the beams, shown in Fig.~3(b), are numbered $1$ to $9$ from left to right. Each beam works as an artificial mechanical atom similar to the theory. We have designed the beams with spatially symmetric via left-right exchange, although in the actual device, slight differences exist due to fabrication imperfection. This does not damage the performance of generating unidirectional transduction. The natural frequencies of the beams range from $1.1~$MHz to $1.4~$MHz, with quality factors around $10,000$ (see SM for detailed parameters). The typical character size of the beams is of the order of tenths of miro-meters. The excitation and detection of the beams are realized via the forces generated by standard magneto-drives \cite{SAA1999} under a magnetic field of $2~$T. The experiment is carried out at $4~$K in vacuum. It is noted that magneto-drive is actually not necessary for our system, as, for example, pure electrical based method is compatible to the double clamped beam similar to our system \cite{NP2008}.

The parametric coupling between two adjacent beams is realized via the electric-static force generated by, similar to our previous work  \cite{PRL2013}, the sum of the applied d.c.\ and a.c.\ voltages between the two beams, with the frequency of a.c.\ voltage being the pump frequency $\omega_{i}^{p}$.  The modulation $\delta_{i}(t)$ is the key for the unidirectional transduction. In practical realization, we employed squared wave modulation instead of the sine wave modulation that used in our theoretical model, as shown in Fig.~3(c). This does not change the unidirectional character but can be simply implemented by gating the continuous wave with pulses (see comparison in SM).

Fig.~4(a)-(c) plots the characteristic frequency response of the system.  When only two adjacent beams ($i$ and $i+1$) are coupled using continuous wave coupling $\delta_{i} (t)=1$ and the pump voltage $V_{i}(t)$ is strong enough, the frequency response (Fig.~4(a)) of the oscillator $i$ exhibits a normal mode splitting $\Delta \omega_{i}$, from which we can estimate the coupling strength as $\eta_{i} = 4\Delta \omega_{i}\sqrt{k_{i}k_{i+1}/\omega_{i}\omega_{i+1}} $ (see SM). Similarly, the frequency responses with 3-4 adjacent beams being coupled together are plotted in Fig.~4(b) and (c) (using the same coupling strengths as the case when only two beams are coupled), showing the merging of splitting.

Fig.~4(d)-(f) plots the unidirectional transduction of the lattice with $N=7$ and its programmability. As expected, when negative delay is used, the forward transmitted signal is suppressed while the backward propagation is allowed (Fig.~4(d)), with an isolation as high as $24~$dB being achieved. In contrast, positive delay leads to the suppression of the backward transmitted signal but allows forward propagation (Fig.~4(e)). When continuous coupling is employed, bi-directional transmissions are observed (Fig.~4(f)), showing the feature of a common transmission line. Therefore, the system becomes programmable by setting different coupling voltage $V_{i}(t)$ topology. Numerical simulations show that larger lattice number ($N=9$) leads not only to higher isolation, but also to stronger decay of energy to the environment. Then stronger couplings, i.e., higher voltages, are required to make the transduction practical, which is not feasible in our current system. The data for $N=9$ and $N =5$ are provided in support online material.

In conclusion, we have proposed and experimentally implemented a radio frequency unidirectional transduction in a parametrically coupled 1D mechanical artificial lattice. The implemented device has already worked as an on-chip and magnetic-free high performance isolator with high controllability. Furthermore,  by simply changing the control topology, the system can be explored for other functions such as quantum limited parametric amplification \cite{RMP2010}. When extended to 2D lattice, the system described here could be utilized to realize artificial gauge field \cite{NPHO2012} and used to build programmable on-chip radio frequency topological device \cite{NP2012}.\\

We thank Sixia Yu, Hao Zeng, Fei Xue and Zhujing Xu for helpful discussions. We thank USTC Center for Micro and Nanoscale Research and Fabrication and Suzhou Institute of Nano-Tech and Nano-Bionics for nano-fabricating supports and Supercomputing center of USTC for calculation supports. This work was supported by the 973 Program (Grant No.~2013CB921800), the NNS-FC (Grant Nos. 11227901, 91021005, 11104262, 31470835, 21233007, 21303175, 21322305, 11374305, 11304356 and 11274299), the Strategic Priority Research Program (B) of the CAS (Grant Nos. XDB01030400 and XDB01020000).

\clearpage

\begin{figure}[h]
\includegraphics[width=\linewidth]{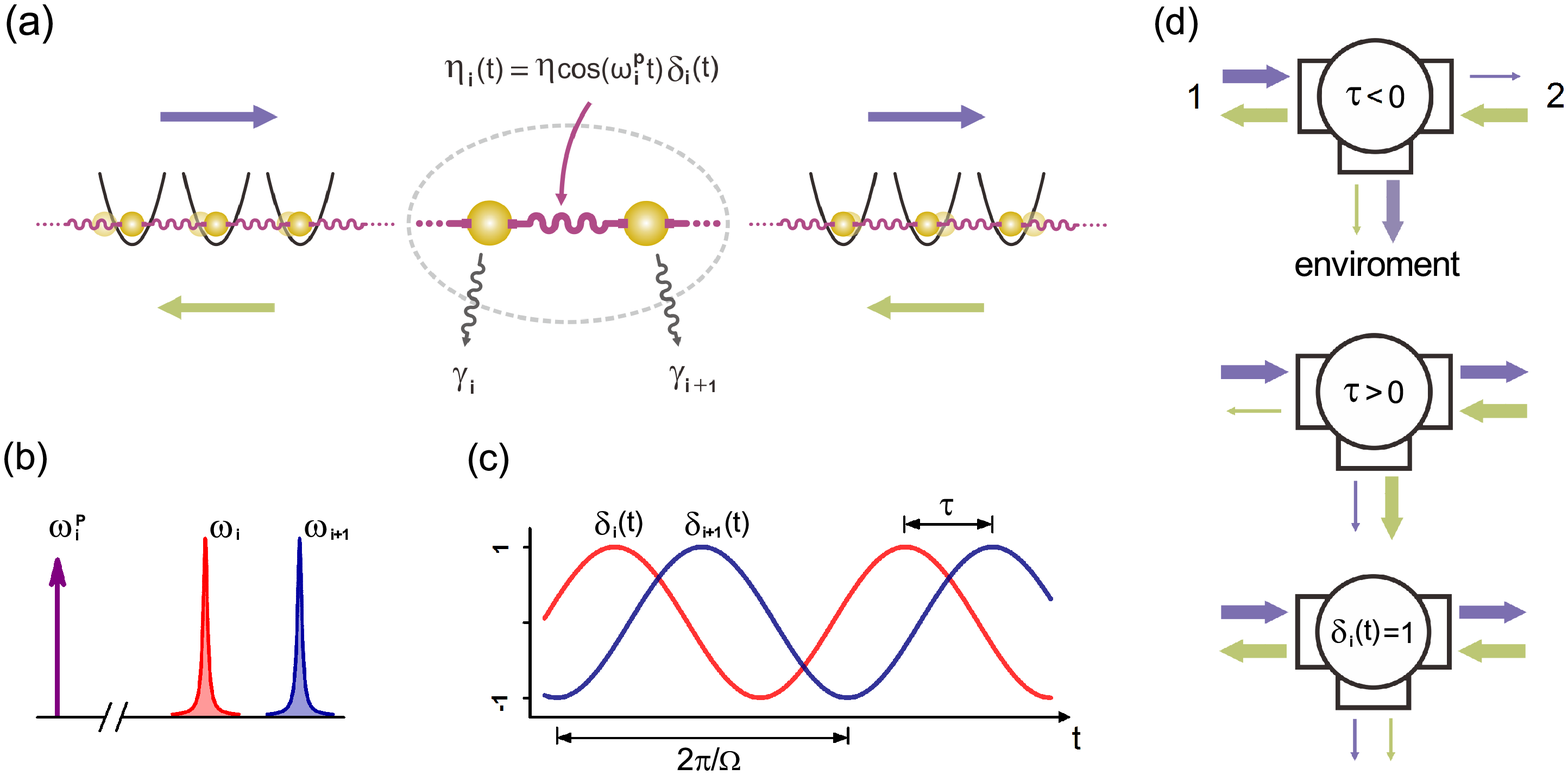}
\caption{Concept of the mechanical artificial lattice.
(A) A chain of harmonic oscillators with adjacent ones being  parametrically coupled, with coupling strength $\eta_{i}\cos(\omega_{i}^p t)\delta_{i}(t)$ between oscillator $i$ and $i+1$, and each oscillator has a decay rate $\gamma_{i}$ due to its interaction with the environment. Coherent oscillation wave propagates forward (purple arrow) or backward (yellow arrow) unidirectionally in such a lattice by designing coupling topology.
(B) The parametric coupling frequency $\omega_{i}^p$ is set to satisfy  the frequency conversion condition $\omega_{i}^p = |\omega_{i}- \omega_{i+1}|$.
(C) The periodic modulation $\delta_{i}(t)= \cos(\Omega t -i \Theta )$ can be taken as a delay $\tau= \Theta/\Omega$ of oscillator $(i+1)$ relative to oscillator $i$.
(D) The results of such a lattice can be expressed as a three port device, with one port connected to environment. For positive delay ($\tau>0$), forward coherent signal input at port 1 could transport to port 2 with only weak decay into environment. While for negative delay, only backward signal passes. For continuous coupling, i.e., $\theta_{i}(t)=1$, the device is bi-directional in that both forward and backward signals.
}
\label{structure}
\end{figure}

\clearpage

\begin{figure}[h]
\includegraphics[width=0.95\textwidth]{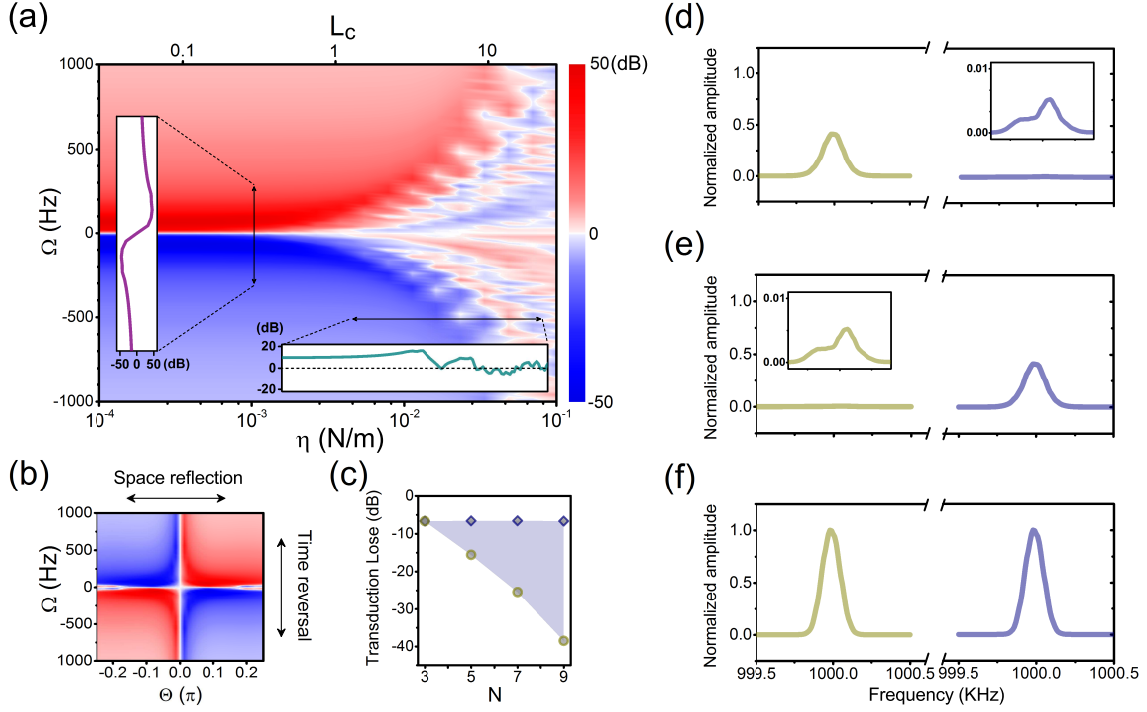}
\label{structure}
\caption{
Numerical simulation of the model.
(A) Dependence of the isolation of forward transduction ${{I}_{\rm for}}$ as a function of coupling strength $\eta$ (or alternatively, the corresponding normalized coherence length $L_c$) and the modulation frequency $\Omega$. The modulation takes the form $\delta_{i}(t) = \cos(\Omega t - i \Theta  )$, with $\Theta = \pi/8 $ and the lattice number $N = 9$. Insets are the profiles indicated by the arrows.
(B) Same as A but as a function of $\Theta$ and $\Omega$, with $\eta = 0.001~$N/m.
(C) Transduction losses for forward (blue points) and backward (yellow points) propagations as a function of lattice number $N$. Positive delays are employed and the parameters are set for each $N$ so that the forward transduction loss is always around $-6~$dB and at the same time, the isolation is optimized (indicated by light blue regime).
(D) The system's frequency responses to forward (left panel) and backward (right panel) transductions, with inset the zooming in of the panel. The modulation is $\delta_{i}(t) = \cos(\Omega t - i \Theta )$ with $\Omega=50~$Hz, $\Theta= \pi/8$ and $\eta=0.001~$N/m. The responses are normalized to the peak values when $\delta_{i}(t)=1$.
(E) Same as (D) but for the time reversal case, i.e., $\Omega \rightarrow -\Omega$.
(F) Same as (D) but under continuous coupling $\delta_{i}(t)=1$. The peak values equal to one due to normalization.
}
\end{figure}

\clearpage

\begin{figure}[h]
\includegraphics[width=\textwidth]{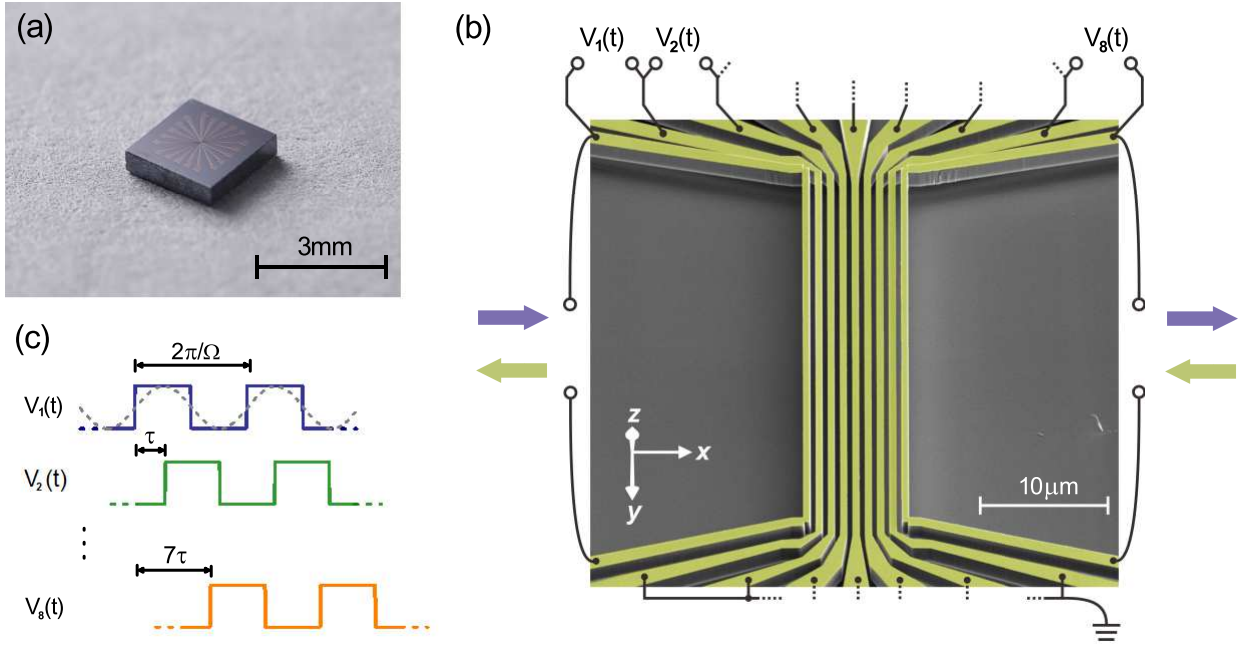}
\caption{
Experimental device and scheme.
(A)  Photograph of the device chip, whose size is about $3~$mm$\times3~$mm. The mechanical artificial lattice is located in the middle, while the effective device is less than $100~\mu$m, or a millionth of the wave length of the electromagnetic field of the same characteristic frequency.
(B) Scanning electron microscopy of the artificial lattice in false color. The oscillators are clamped silicon beams with a thin layer of gold deposited on top of them.
The first oscillation mode of each beam vibrates along $x$ axis. The magnetic field is applied along $z$ axis for excitation (input) and detection (output), indicated by leftmost and rightmost arrows. Parametric couplings which are used to programme the functionality of the lattice are realized by voltage $V_{i}(t)$ applied between adjacent beams.
(C) The pulse scheme for realization. Parametric coupling between beams $i$ and $i+1$ is realized by a continuous pumping wave $\eta_i \cos(\omega_{i}^p t)$ modulated by the plotted squared wave. The repetition frequencies for all channels are the same $\Omega$, while a delay $\tau$ is set between adjacent channels.
}
\label{circuit}
\end{figure}

\clearpage

\begin{figure}[h]
\includegraphics[width=0.9\textwidth]{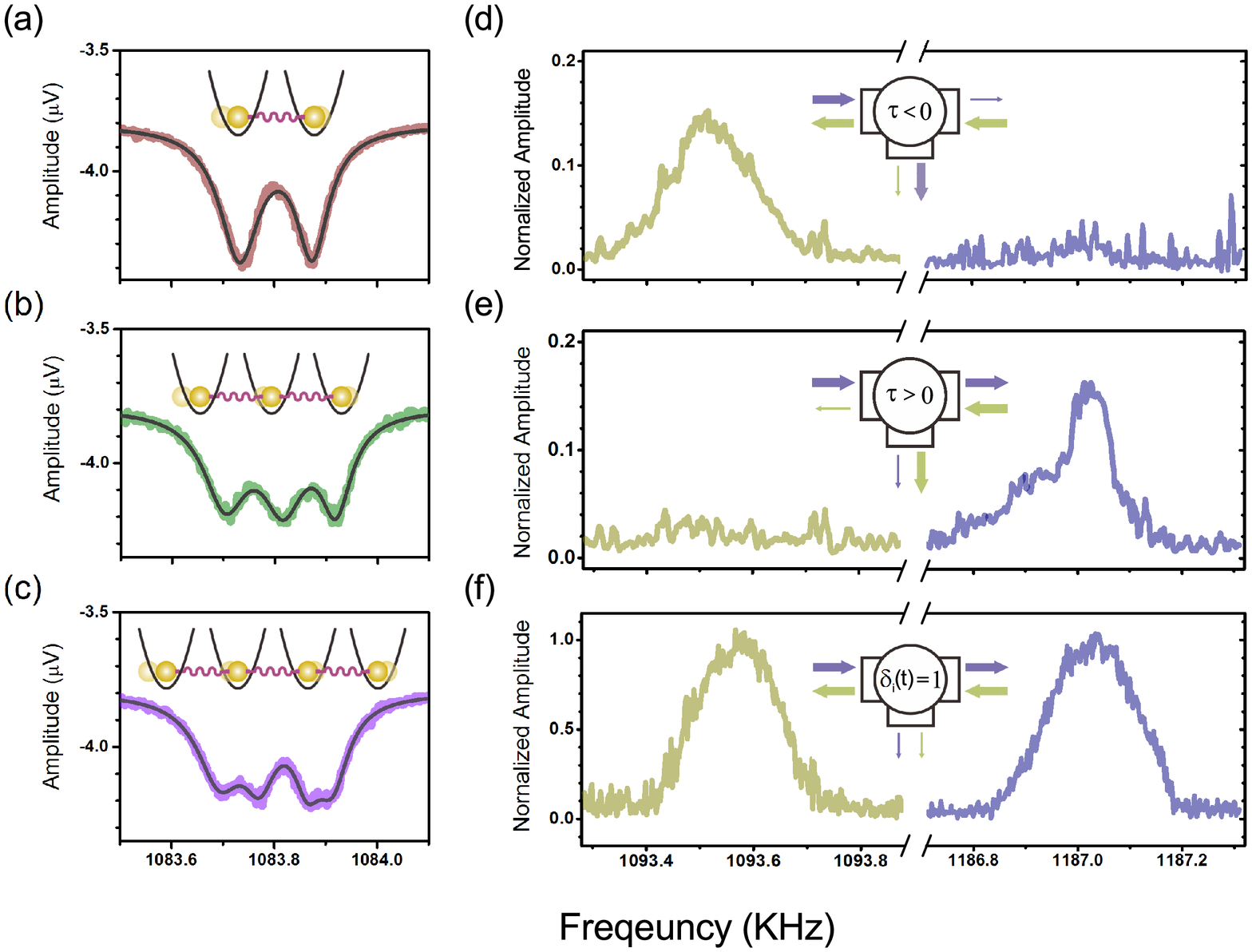}
\caption{Results of programmable unidirectionality transduction.
(A) A represented frequency response when only two adjacent beams are coupled by continuous pump $\delta_{i} (t) =1 $. The splitting $\Delta \omega_{i}$ is used to calibrate the coupling strength (data from beam No.~7 with input to the coupled beam No.~6). Dots are measured data and solid curve is Lorentz fitting to guide eyes.(B,C) same as (A) under the same pump strength but with three beams (No.~5 , No.~6 and No.~7) and four beams (No.~4 , No.~5 No.~6 and No.~7) coupled, respectively.
(D) The transduction of the lattice with $N = 7$ under time dependent modulation with negative delay, parameter used are $\Theta = \pi/6$ , $\Omega = 50~$Hz and $\eta_{i} \approx $ $0.001~$N/m. Left is the backward result with excitation applied on beam No.~3 and amplitude measured on No.~9, normalized to response of continuous pump. (E) The same as (D) but with positive delay. (F) The same as (D) but with continuous drive $\delta_{i} (t) =1 $.
}
\label{tomo}
\end{figure}

\clearpage


\begin{thebibliography}{99}
\bibitem{Landau1986em} L. D. Landau, E. M. Lifshitz and L. P. Pitaevskii, \textit{Electrodynamics of Continuous Media Electrodynamics of Continuous Media 2nd edition} (Elsevier, Singapore, 2007).

\bibitem{Landau1986fm} L. D. Landau, and E. M. Lifshitz,  \textit{Fluid Mechanics 2nd edition} (Elsevier, Singapore, 2007).

\bibitem{NPHO2011}L. Bi, J. Hu, P. Jiang, D. H. Kim, G. F. Dionne, L. C. Kimerling, and C. A. Ross, {Nat. Photonics} \textbf{5}, 758 (2011).

\bibitem{TE1965} S. Tanaka,  N. Shimimura, and K. Ohtake, Proc. IEEE \textbf{53}, 260 (1965).

\bibitem{APL2011} T. Kodera, D. L. Sounas, and C.Caloz, Appl. Phys. Lett. \textbf{99}, 03114 (2011).


\bibitem{JAP1994} M.  Scalora,  J. P. Dowling,  C. M. Bowden, and M. J. Bloemer, {J. Appl. Phys.} \textbf{76}, 2023 (1994).

\bibitem{JOS2002} S. F. Mingaleev, and Y. S. Kivshar, {J. Opt. Soc. Am. B} \textbf{19}, 2241 (2002).

\bibitem{NM2005} B. Liang, X. S. Guo, J. Tu, D. Zhang, and J. C. Cheng   {Nat. Material} \textbf{9}, 989 (2005).

\bibitem{SC2012} L. Fan, J. Wang, L. T. Varghese, H. Shen, B. Niu, Y. Xuan, A. M. Weiner, and M. Qi, {Science} \textbf{335}, 447 (2012).

\bibitem{NP2015}J. Kim, M. C. Kuzyk, K. Han, H. Wang, and G. Bahl, {Nat. Phys.} \textbf{11}, 275 (2015).


\bibitem{NPHO2009} Z. Yu, and S. Fan, {Nat. Photonics} \textbf{3}, 91 (2009).

\bibitem{NP2011} A. Kamal, J. Clarke, and M. H. Devoret, {Nat. Phys.} \textbf{7}, 311 (2011).

\bibitem{NPHO2014} L. D. Tzuang, K. Fang, P. Nussenzveig, S. Fan, and M. Lipson, {Nat. Photonics} \textbf{8}, 701 (2014).

\bibitem{PRX2015} K. M. Sliwa, M. Hatridge, A. Narla, S. Shankar, L. Frunzio, R. J. Schoelkopf, and M. H. Devoret, {Phys. Rev. X } \textbf{5}, 041020 (2015).

\bibitem{NPHO2014_2}L. Chang, X. Jiang, S. Hua, C. Yang, J. Wen, L. Jiang, G. Li, G. Wang, and M. Xiao, {Nat. Photonics} \textbf{8}, 524 (2014).

\bibitem{NP2014}B. Peng, \c{S}. K. \"Ozdemir, F. Lei, F. Monifi, M. Gianfreda, G. L. Long, S. Fan, F. Nori, C. M. Bender, and L. Yang, {Nat. Phys.} \textbf{8}, 394 (2014).


\bibitem{NC2013} D. L. Sounas, C. Caloz, and A. Al\`{u}, {Nat. Commun.} \textbf{4}, 2047 (2013).

\bibitem{SC2014}R. Fleury, D. L. Sounas, C. F. Sieck, M. R. Haberman, and A. Al\`{u}, {Science} \textbf{343}, 516 (2014).

\bibitem{NP2014_2} N. A. Estep, D. L. Sounas, J. Soric, and A. Al\`{u}, {Nat. Phys.} \textbf{10}, 923 (2014).


\bibitem{NA2009} Z. Wang, Y. Chong, J. D. Joannopoulos, and Marin Solja\v{c}i\'{c}, {Nature (London)} \textbf{461}, 772 (2009).

\bibitem{NM2012}A. B. Khanikaev, S. Hossein Mousavi, W. K. Tse, M. Kargarian, A. H. MacDonald, and G. Shvets, {Nat. Material} \textbf{12}, 233 (2012).

\bibitem{SC2015} R. Susstrunk, and S. D. Huber, {Science} \textbf{349}, 47 (2015).

\bibitem{NA2002} M. Greiner, O. Mandel, T. Esslinger, T. W. H\`{a}nsch, and I. Bloch, {Nature (London)} \textbf{415}, 39 (2002).

\bibitem{SC2011}A. Singha, M. Gibertini, B. Karmakar, S. Yuan, M. Polini, G. Vignale, M. I. Katsnelson, A. Pinczuk, L. N. Pfeiffer, K. W. West, and V. Pellegrini, {Science} \textbf{332}, 1176 (2011).

\bibitem{NA2013}C. R. Dean, L. Wang, P. Maher, C. Forsythe, F. Ghahari, Y. Gao, J. Katoch, M. Ishigami, P. Moon, M. Koshino, T. Taniguchi, K. Watanabe, K. L. Shepard, J. Hone, and P. Kim, {Nature (London)} \textbf{497}, 598 (2013).

\bibitem{SC2000} H. G. Craighead, {Science} \textbf{290}, 1532 (2000).

\bibitem{RSI2005} K. L. Ekinci, and M. L. Roukes, {Rev. Sci. Instrum.} \textbf{76}, 061101 (2005).


\bibitem{PRL2013}P. Huang, P. Wang, J. Zhou, Z. Wang, C. Ju, Z. Wang, Y. Shen, C. Duan, and J. Du, {Phys. Rev. Lett.} \textbf{110}, 227202 (2013).

\bibitem{SAA1999} A. Cleland,  and M. Roukes, {Sensors and Actuators A} \textbf{72}, 256(1999).

\bibitem{NP2008} C. A. Regal, J. D. Teufel, and K. W. Lehnert, {Nat. Phys.} \textbf{4}, 555 (2008).

\bibitem{RMP2010}A. A. Clerk, M. H. Devoret, S. M. Girvin, F. Marquardt, and R. J. Schoelkopf, {Rev. Mod. Phys.} \textbf{82}, 1155 (2010)


\bibitem{NPHO2012} K. Fang, Z. Yu, and S. Fan, {Nat. Photonics} \textbf{6}, 782 (2012)

\bibitem{NP2012}M. Hafezi, E. A. Demler, M. D. Lukin, and J. M. Taylor, {Nat. Phys.} \textbf{7}, 907 (2012)


\end{thebibliography}
\end{document}